\documentclass[fp,twocolumn]{jpsj3}
\usepackage{graphicx}
\usepackage{txfonts}
\usepackage{bm}
\usepackage[ruled,vlined]{algorithm2e}
\usepackage{xcolor}
\usepackage[normalem]{ulem}
\usepackage{epsfig}
\usepackage{amsmath}

\title{Phase Transition of Hard Disk Systems with Vicsek-type Interactions}

\author{Nobuaki Murase$^1$\thanks{n.murase.656@stn.nitech.ac.jp}, Masaharu Isobe$^1$\thanks{isobe@nitech.ac.jp}}
\inst{$^1$Graduate School of Engineering, Nagoya Institute of Technology, 466-8555, Japan}

\abst{The phase diagram of self-propelled hard disk systems with Vicsek-type alignment interactions was investigated by event-driven molecular dynamics simulations.
The model incorporates two competing order parameters: the polar order-disorder transition associated with collective velocity alignment (Vicsek model) and the orientational order arising from solid-fluid transitions (Alder transition) induced by excluded volume effects.
The incompressibility of hard disks suppresses motility-induced phase separation at high packing fractions.
Distinctive fluctuations were observed near the transition point, accompanied by anomalous shifts in the transition point as functions of noise intensity and packing fraction.
Analysis of local configurational parameters---specifically, orientational order and circularity of free volume---provides insight into the microscopic origins of these anomalous phase transition shifts.}

\begin{document}
\maketitle 

\section{Introduction}
\label{sec:1}
Active matter systems, composed of self-propelled elements, have garnered significant interest in the field of non-equilibrium physics due to their ability to exhibit collective motion and phase transitions~\cite{pismen_2021,ramaswamy_2010,marchetti_2013,cates_2015,Peruani_2011}.
The Vicsek model (VM) is a seminal minimal model that captures the essential features of active matter systems, such as the emergence of ordered collective motion from local interactions~\cite{vicsek_1995, vicsek_2012}.
The VM has been extensively studied and has been shown to belong to a novel universality class in non-equilibrium systems~\cite{Ginelli_2016, Chate_2020}.
However, the original VM treats particles as point-like objects without considering excluded volume effects, which are crucial in real-world systems.
The presence of finite-sized particles and their excluded volume interactions can significantly influence the collective behavior and phase transitions in active matter systems~\cite{chat_2008,Peruani_2011}.

Recent studies have investigated the phase diagram of self-propelled particles, including the shift in the freezing transition, by introducing various interactions between elements in two dimensions~\cite{bialke_2012, Digregorio_2018,Weber_2012,render2013,singh_2016,cugl_2017,digre2018,pali2020,loewe2020,pasu2020,digre2020,shi2023}.
Meanwhile freezing/melting transitions in hard sphere/disk systems have been a central topic in the history of molecular simulations and liquid theory in statistical mechanics~\cite{chandler_1987, krauth_2006, frenkel_2002, hansen_2013, allen_2017, isobe_2016}, the change in the packing fraction induces a breakdown of the symmetry of the orientation and position of the particles, leading to a transition from an ordered crystal (solid) phase to a disordered (fluid) phase~\cite{alder_1957,wood_1957,alder_1962,bernard_2011,engel_2013,isobe_2015}, known as the Alder transition.
This order-disorder transition originates from the geometric properties of the particles, with the packing fraction (density) serving as the control parameter.

In this work, we introduce a modified Vicsek model that incorporates hard disk repulsive forces to account for excluded volume effects.
We aim to elucidate the impact of self-propulsion on the freezing/melting transition in hard disk systems with respect to free volume (entropic) contributions,~\cite{mugita2024} focusing on the violation of momentum conservation at the particle level, a characteristic feature of non-equilibrium systems.
By performing efficient event-driven molecular dynamics (EDMD) simulations~\cite{alder_1959, rapaport_1980, isobe_1999, rapaport_2004} in a dense two-dimensional hard disk system, we investigate the competition between the frequency of repulsive elastic collisions (particle inertia) and VM interactions in the order parameters.
Our study reveals a shift in the phase transition point in the VM with hard disks and provides insights into the microscopic origins of this shift by analyzing polar and orientational order parameters, local structure (free volume, and local circularity), the incompressibility of crystal states of highly dense systems.
Understanding the complex interplay between self-propulsion, excluded volume effects, and the crystallization transition is crucial for advancing our knowledge of active matter systems and their non-equilibrium behavior.

The remainder of this paper is organized as follows. Section 2 describes the model and order parameters. Section 3 presents the simulation results for the polar order parameter (Sec. 3.1), orientational order parameter (Sec. 3.2), and local structure analysis (Sec. 3.3). Section 4 provides concluding remarks. Technical details are provided in Appendices A and B.

\section{Model and Order Parameters}
\label{sec:2-0}

The original VM is a minimal model describing collective motion, typically assuming that point particles move at a constant speed $v_0$ and align with the average direction of their neighbors~\cite{vicsek_1995, vicsek_2012}.
In contrast, our model incorporates excluded volume effects using hard disks.
Consequently, the speed of each particle $v_i(t)$ is not constant but fluctuates dynamically due to elastic collisions. After a sufficient simulation time, the velocity distribution relaxes to the Maxwell--Boltzmann distribution.
To account for this, we adopt a modified interaction rule where the new direction is determined by the vector sum of the velocities of neighboring particles, explicitly depending on their varying speeds $v_j$, rather than just their orientations.
The velocity vectors of all elements in the system are updated at every interval time $\Delta t$.
The update rule is given by:

\begin{eqnarray}
\Theta_i (t+\Delta t) & =& \arg{\left( \sum_j v_j \exp{({\rm {\bf i}} \Theta_j (t)) }\right)} + \delta \Theta_i (t), \label{eqn:vicsek1} \\
\bm{v}_i (t+\Delta t) & = & v_i (t) \bm{\hat{e}}_{\Theta_i (t+\Delta t)},
\label{eqn:vicsek2}
\end{eqnarray}

\noindent
where ${\bm v}_i (t)$ is the velocity vector of element $i$ at time $t$, and $v_i (t)=|\bm{v}_i (t)|$ and $\Theta_i (t)$ are its magnitude and direction, respectively.
The summation in eq.~(\ref{eqn:vicsek1}) is taken over all neighbors $j$ (including particle $i$ itself) within the cutoff radius $r_c$.
The second term represents the fluctuation, where a uniformly distributed noise $\delta \Theta$ is chosen from the interval $[-\eta/2, \eta/2]$.
$\bm{\hat{e}}_{\Theta_i (t)}$ in eq.~(\ref{eqn:vicsek2}) is the unit vector in the $\Theta_i (t)$ direction.

Hard disks with a radius of $\sigma$ (number of particles $N=4096$) are placed inside a rectangular box $L_y \times L_x$, where the aspect ratio is $L_y / L_x = \sqrt{3}/2$, with periodic boundary conditions.
This aspect ratio realizes a close packed two-dimensional crystal structure (triangular lattice) without defects.
The packing fraction is defined as $\nu= \pi N \sigma^2/ (L_x L_y)$.
In equilibrium hard disk systems, the solid-fluid phase transition (Alder transition) occurs above $\nu_c \sim 0.70$ \cite{note_hexatic}.
We introduced this Vicsek-type interaction with the interval time $\Delta t$ to hard disk systems to investigate the excluded volume effect through efficient EDMD~\cite{alder_1959,rapaport_1980,isobe_1999}.
In EDMD, the time evolution algorithm is event-based.
Therefore, the number of elastic collisions within the time scale $\Delta t$ directly affects the dynamics of the system.
To investigate this competition, we introduce the parameter $\Delta t^* = \Delta t/\tau \ (=1 \sim 100)$, in units of the non-dimensional averaged collision time $\tau$ in the hard disk systems at $\nu=0.720$ in the equilibrium state.
The cutoff length of the VM interaction is set at $r_c^*=r_c/(2\sigma) = 1.556$, which roughly corresponds to the first minimum of radial distribution functions in the equilibrium hard disk system at $\nu=0.720$.

To characterize the phase for analyzing global properties in the hard disk system with Vicsek interaction, we focus on two intrinsic order parameters.
One is the polar order parameter $\psi$ to detect the collective flow of elements, which has been well studied in the original VM~\cite{vicsek_2012}.

\begin{equation}
\psi = \frac{\left| \sum_{i=1}^N {\bm v}_i\right|}{\sum_{i=1}^N \left|{\bm v}_i \right|}.
\end{equation}
At $\eta = 2.0 \pi$ in the original VM, the elements are disturbed by a Langevin-type heat bath at the equilibrium state $\psi = 0$ (disorder phase), since all directions are chosen by noise equivalently.
On the other hand, at $\eta = 0.0$, the elements in the whole system flow in the same direction through PBCs, $\psi = 1$ (ordered phase).

The other is the (six-fold symmetry) hexatic orientational order parameter in dense hard disk systems,
\begin{equation}
\phi_6^i = \frac{1}{N_j} \sum_{j=1}^{N_j} \exp{(6{\bf i} \theta_j^i)},
\end{equation}

\noindent
for each element $i$, where $j$ is the nearest neighbor element within the cut-off radius $r_c^* $ (equivalent values to those of VM interactions), $N_j$ is its total number, and $\theta_j^i$ is the angle of $j$ from the reference axis (e.g. $x$ axis) through the center of element $i$.
Using $\phi_6^i $, the local and global orientation order parameters of the whole system are defined as,

\begin{eqnarray}
\Phi_6^{\rm L} & = & \frac{1}{N} \sum_{i=1}^N \left|
\phi_6^i \right|, \\
\Phi_6^{\rm G} & = & \left| \frac{1}{N} \sum_{i=1}^N \phi_6^i \right|,
\end{eqnarray}

respectively. 
In the closed packing state in the hard disk systems $\nu \sim 0.906$, the hexatic orientational order parameter becomes unity $(\Phi_6^L=\Phi_6^G=1) $, where the system is in the perfect crystal state.
In contrast, in configurations with non-six-fold symmetry (such as four-fold symmetry in the crystal or liquid states), $\Phi_6^{\rm L} =\Phi_6^{\rm G}=0$.

For the analysis of the local structure of the elements, the free volume, the free surface area surrounding each element $i$ and the defects are investigated ~\cite{mugita2024}.
Since we consider the two dimensional hard disk systems, the free volume $v_f^i$ of element $i$ is defined as the area where the center of the tagged element can move under the condition that the surrounding elements are fixed at a certain configuration, and the free surface $s_f^i$ is the length of the surrounding curve (perimeter) of the free volume for element $i$.

In hard disk systems, the local circularity measure of each disk $c^*_i$ and that of the entire system $\cal C^*$ can be defined by Eqs. ~(\ref{eqn:circularity0}) and (\ref{eqn:circularity}), respectively, where $c^*_i$ in two dimensions can be defined as the ratio between area and perimeter of free volume rigorously estimated numerically by NELF-A algorithm~\cite{mugita2025}.
\begin{eqnarray}
c^*_i & = &  \frac{{s_f^i}^2}{4\pi {v_f^i}}, \label{eqn:circularity0}\\
\cal C^* & = & \frac{1}{N}\sum_{i=1}^N c_i^*,
\label{eqn:circularity}
\end{eqnarray}
where $c_i^* $ takes a value of 1 for circles and values greater than 1 for non-circular shapes.

\section{Results}
\label{sec:3-0}

\subsection{Polar Order Parameter}

\begin{figure}
\includegraphics[width=\linewidth]{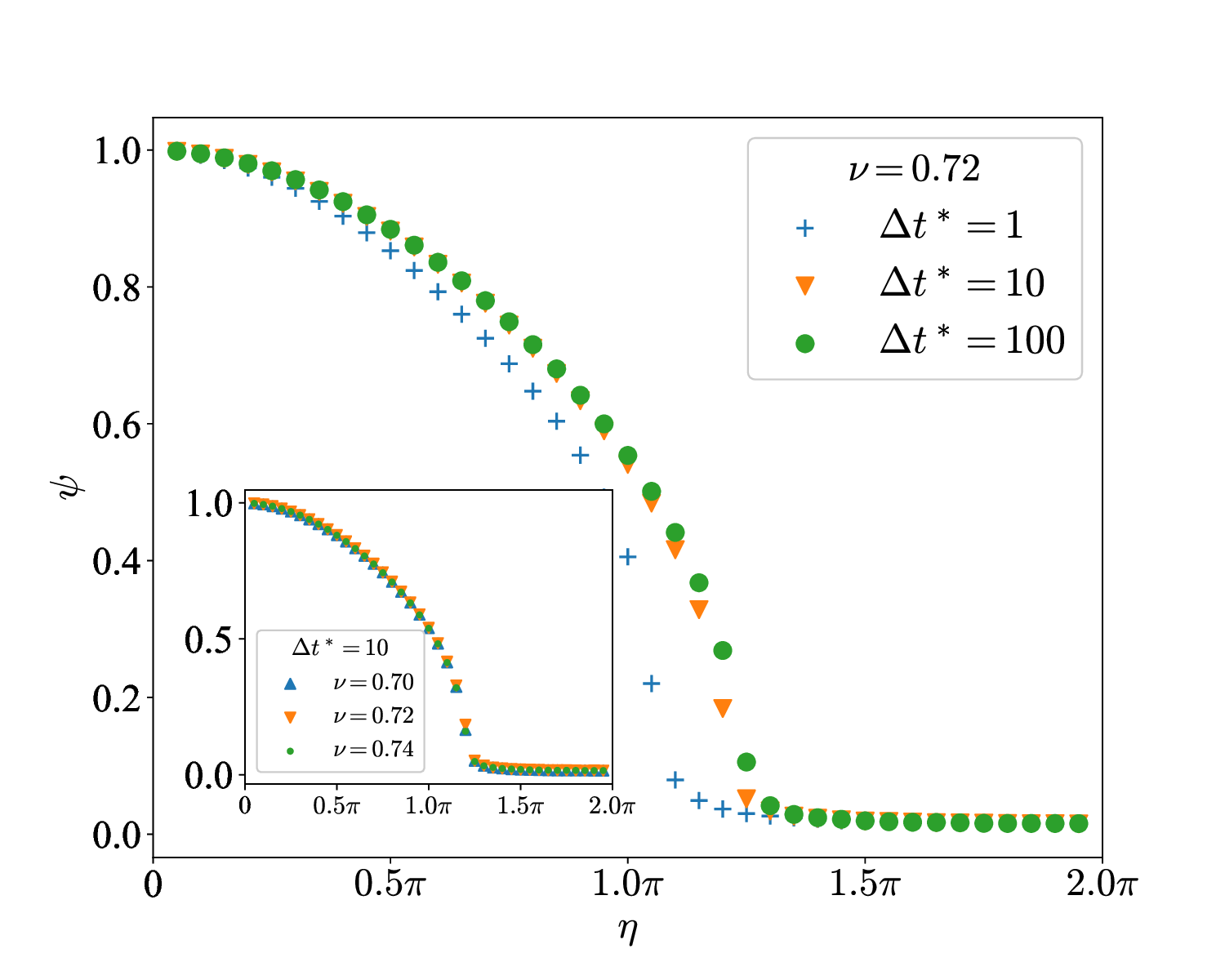}
\caption{(Color online) The polar order parameter $\psi$ as a function of noise $\eta$ in densely packed hard disk systems with $\nu=0.72$, for various values of the Vicsek interaction time interval $\Delta t^*$.
The inset shows the packing fraction $\nu$ dependence of $\psi$ at a fixed $\Delta t^*=10$.}
\label{fig:1}
\end{figure}

We observed that an order-disorder phase transition occurs in this systems around $\eta \approx 1.2\pi$.
As shown in Fig.~1, we also found that the behavior of $\psi$ with respect to $\eta$ is robust for packing fractions $\nu$ above the Alder transition point ($\nu \approx 0.70$).
These results imply that the number of neighbors involved in the Vicsek interaction does not change significantly due to the incompressible nature of densely packed hard disks.
If the frequency of the Vicsek interaction increases (i.e., smaller $\Delta t^*$), a shift in the transition point of $\psi$ can be observed.
This suggests an intrinsic competition between elastic collisions within the local cage and the Vicsek interaction.

In densely packed hard disk systems with Vicsek interactions, $\psi$ rapidly decreases as the noise $\eta$ increases, which is characteristic of the VM.
We observed that an order-disorder phase transition occurs in this system around $\eta \approx 1.2\pi$.
Regarding the nature of this transition in the VM with the cutoff method, it is widely recognized as a discontinuous (first-order) transition in the thermodynamic limit~\cite{Ginelli_2016, Chate_2020, chat_2008}.
However, since our system size is relatively small, we obtain results similar to the continuous transition behavior reported in the early work of Vicsek et al.~\cite{vicsek_1995}.
While first-order characteristics may emerge with an increased number of particles, this study focuses on the description of the ordered and disordered phases of velocity and the state near the transition point;
thus, the rigorous determination of the transition order is outside the scope of this paper.

\subsection{Orientational Order Parameter}
\begin{figure}
\includegraphics[width=\linewidth]{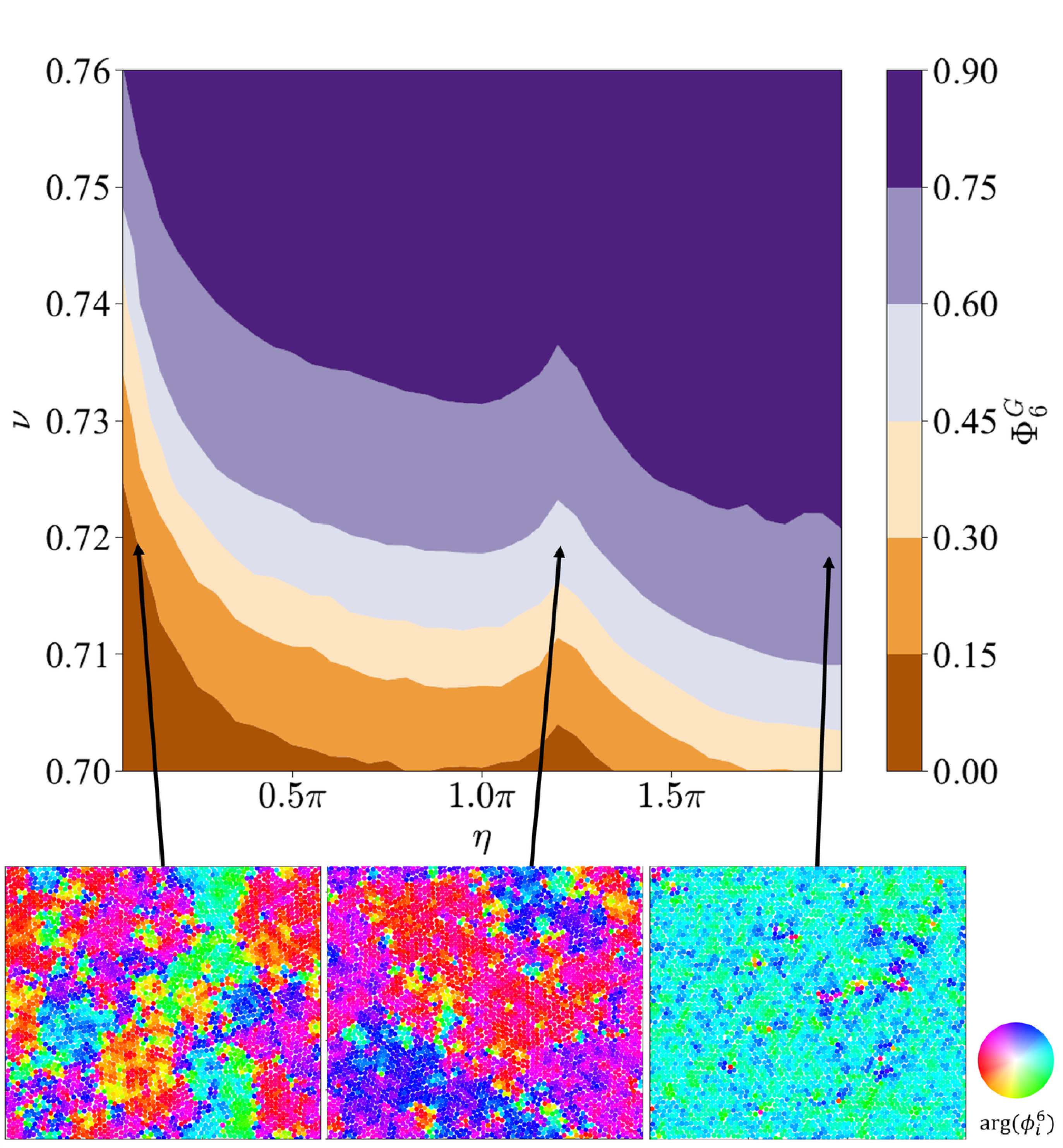}
\bigskip
\caption{(Color online) Contour map of the global orientational order parameter $\Phi_6^{\rm G}$ in the $\nu - \eta$ plane at $(N, \Delta t^* ) = (4096, 10)$.
The snapshots color-coded by orientation direction $arg(\phi_6^i)$, at $\nu = 0.72$ for $\eta =0.1, 1.2\pi$, and $1.9\pi$ (from left to right)}.
\label{fig:2}
\end{figure}

Figure~\ref{fig:2} shows a contour map of the global orientational order parameter $\Phi_6^{\rm G}$ in the $\nu - \eta$ plane for a fixed $\Delta t^* = 10$.
In hard disk systems, $\Phi_6^{\rm G}$ allows us to identify the transition from a six-fold symmetric crystalline (ordered) phase to a liquid (disordered) phase.
As the packing fraction $\nu$ increases, $\Phi_6^{\rm G}$ generally increases, and the system transitions into a crystalline state, regardless of the noise value $\eta$.
At lower noise levels $\eta$ (a non-equilibrium state), $\Phi_6^{\rm G}$ tends to be lower compared to that at larger $\eta$ for the same $\nu$.
Consequently, the phase boundary for the crystalline state shifts significantly toward higher $\nu$ as $\eta \to 0$.
In addition, a non-trivial cusp can be observed around $\eta \sim 1.2 \pi$.

\begin{figure}
\includegraphics[width=\linewidth]{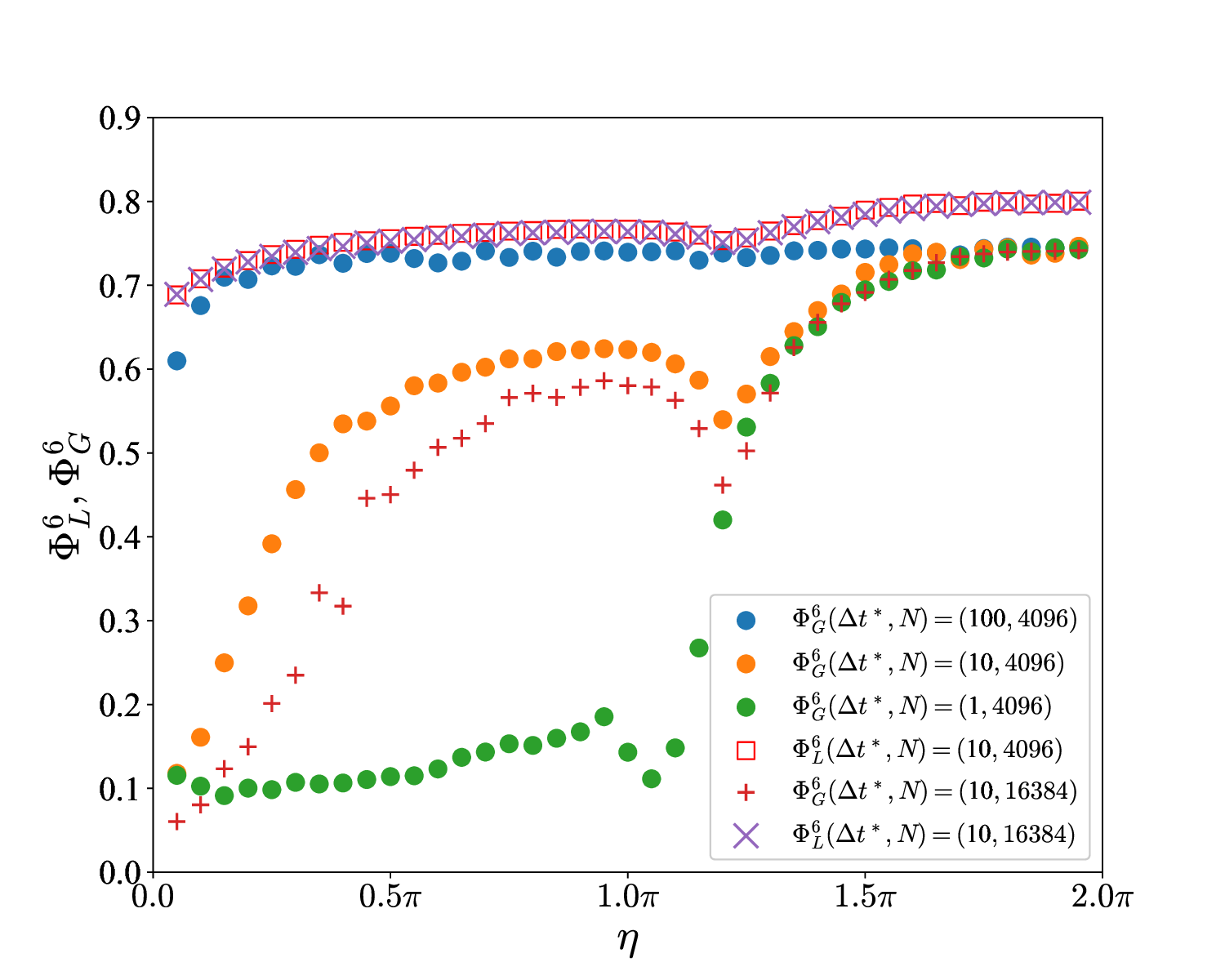}
\caption{(Color online) The global orientational order parameter $\Phi_6^{\rm G}$ and local orientational order parameter $\Phi_6^{\rm L}$ as a function of $\eta$ for different $\Delta t^*$ and $N$ at a fixed packing fraction $\nu = 0.72$.}
\label{fig:3}
\end{figure}

Then, we focus on the dependence of $\Phi_6^{\rm G}$ on $\eta$ at a fixed $\nu=0.72$, while varying the Vicsek interaction interval $\Delta t^*$, as shown in Fig.~\ref{fig:3}.
For the equilibrium-like system ($\eta \approx 2.0 \pi$), $\Phi_6^{\rm G} \approx 0.71$, indicating a crystalline state.
Conversely, for $\eta \leq 1.5\pi$, $\Phi_6^{\rm G}$ decreases. At $\Delta t^*=1$, the system exhibits a liquid-like state around $\eta \approx 1.0\pi$, with $\Phi_6^{\rm G} \approx 0.1$.
A cusp is also observed where $\Phi_6^{\rm G}$ temporarily decreases around $\eta=1.0\pi \sim 1.3\pi$, irrespective of $\Delta t^*$.
In the original VM, the polar order parameter undergoes a phase transition around this range.
In our model, this transition point corresponds to $\eta \approx 1.2\pi$, as shown in Fig.~\ref{fig:1}.
The decrease (cusp) in $\Phi_6^{\rm G}$ is reminiscent of the large fluctuations observed in critical phenomena around a phase transition.

Figure \ref{fig:3} also shows the noise dependence of both the global ($\Phi_6^{\rm G}$) and local ($\Phi_6^{\rm L}$) orientational order parameters at $(\nu, \Delta t^*) = (0.72, 10)$.
In general, both order parameters tend to decrease as $\eta$ decreases.
$\Phi_6^{\rm G}$ vanishes around $\eta \approx 0.0$, while $\Phi_6^{\rm L}$ remains high at $\sim 0.70$.
The significant reduction of $\Phi_6^{\rm G}$ indicates that even though the local orientational order for each hard disk is not strongly affected by $\eta$, the Vicsek interaction disrupts the long-range orientational order throughout the system.
In the low-$\eta$ limit, the system forms polycrystalline clusters, each with a different orientation, separated by grain boundaries (see Fig. ~\ref{fig:2}).
This leads to a decrease in $\Phi_6^{\rm G}$ even when $\Phi_6^{\rm L}$ remains high.
This is a distinct effect of the Vicsek interaction in hard disk systems.
For any $\Delta t^*$, both $\Phi_6^{\rm G}$ and $\Phi_6^{\rm L}$ are approximately 0.71 (a crystalline state) at $\eta \approx 2.0 \pi$, indicating that the system is in a crystal state.
In the case of $\Delta t^* = 1$, $\Phi_6^{\rm G}$ decreases drastically as $\eta$ is decreased, reaching $\Phi_6^{\rm G} \approx 0.1$ around $\eta \approx 1.0 \pi$, which indicates a liquid state.
We confirmed that the behavior of $\Phi_6^{\rm L}$ does not change significantly with system size (see Fig.~\ref{fig:3}).
In contrast, $\Phi_6^{\rm G}$ remains almost unchanged near equilibrium but becomes smaller for small $\eta$ in the larger system.

\begin{figure}
\includegraphics[width=\linewidth]{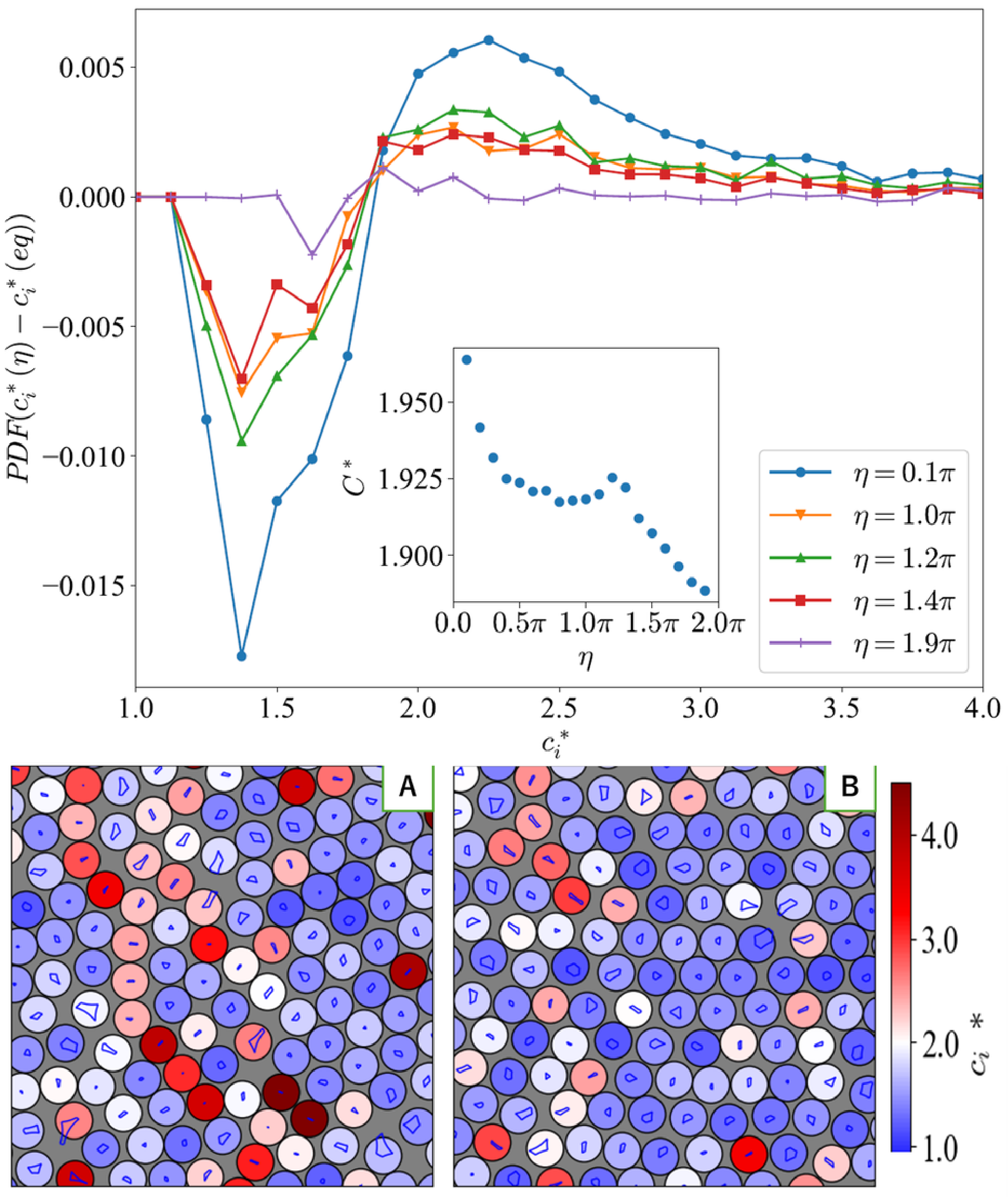}
\caption{(Color online) The probability distribution functions of the local circularity $c_i^*$ based on that of equilibrium are shown for each noise value $\eta$.
The inset shows the total circularity $\cal C$. The snapshots color-coded by $c_i^*$ and free volume shape (drawn in curves) at $\nu = 0.70$ for $\eta = 0.1\pi $ (A) and $1.9\pi $ (B).}
\label{fig:5}
\end{figure}

\subsection{Local Structure}

We now turn to the local structural configurations of neighbors for individual hard disks using the free volume $v_f^i$, the absolute values of the local orientational order $|\phi_6^i$, and the local circularity $c_i^*$.
Figure 4 shows the probability density distribution (PDF) of local circularity $c_i^*$ for each $\eta$ based on equilibrium distributions.
We found a distinct decrease in circularity around $c_i^* \sim 1.4$ and a broad increase around $c_i^* \sim 2.3$, particularly prominent at $\eta = 0.0$ and $\eta = 1.2\pi$.
We also observed that similar trends appeared in distributions of defects (non-six nearest neighbors), free volume $v_f^i$, and the $\phi_6^i$ order parameter.
These transition shifts correlate strongly with non-sixfold (non-circular) symmetry of free volume (indicated by curved lines in the snapshots of Fig. 4).
The inset shows total system circularity, consistent with the polar order transition at intermediate $\eta$ (Fig. 1) and global orientational order at lower $\eta$ (Fig. 2).
Notably, VM interactions induce changes in the PDF of free volume geometry at two distinct circularity ranges, corresponding to rectangular aspect ratios of $\sim$1:2 for $c_i^* \sim 1.4$ and $\sim$1:4 for $c_i^* \sim 2.3$, respectively.
These geometric differences in free volume directly affect the entropy-driven mobility of disks, which is independent of free volume magnitude.
The elongated geometry at $c_i^* \sim 2.3$ facilitates large-displacement hopping events, driving fluidization even when the packing fraction is identical between (A) and (B) in Figure 4. This demonstrates that the geometry (shape) of free volume, not magnitude alone, critically governs collective dynamics and fluidization onset in dense systems.

\section{Concluding Remarks}
\label{sec:4-0}

In this study, we introduced hard disks into the VM to investigate the effects of excluded volume.
The characteristics of the system, including its phase diagram and microscopic origins, were systematically investigated by EDMD simulations, varying the packing fraction $\nu$ and noise intensity $\eta$.
In the original VM with point particles, a reduction in noise $\eta$ aligns the velocity vectors of neighboring particles, quantified by the polar order parameter.
As a result, the system undergoes a phase transition from a disordered, equilibrium-like phase to an ordered phase of collective motion.
On the other hand, in the hard disk system, an increase in $\nu$ causes a solid-liquid phase transition (Alder transition).
This transition is also observed in our model. However, the solid-liquid phase transition boundary shifts to higher packing fractions $\nu$ as the noise $\eta$ is reduced.
Furthermore, a cusp emerges in the orientational order parameter around $\eta \sim 1.2\pi$, near the polar order-disorder transition point.
When the update interval of the Vicsek interaction, $\Delta t^*$, is varied, we observe a clear influence on the overall order of the system, highlighting the competition between the Vicsek interaction frequency and the particle collision rate.
This is one of the distinct differences from the original VM.
An increase in the frequency of the Vicsek interaction also leads to the formation of polycrystalline clusters with different orientation orders and grain boundaries.
Therefore, the global orientational order parameter $\Phi_6^{\rm G}$ of the whole system decreases rapidly even though the local orientational order parameter $\Phi_6^{\rm L}$ remains high.
The analysis of the local structure shows an increase in hard disks with high local circularity $c^*_i$.
There are two factors contributing to high local circularity: either high local packing density among neighbors or an anisotropic free volume shape caused by lattice defects.
We observed that the latter case is dominant at lower $\eta$.
Furthermore, we found that the shape of free volume, rather than its magnitude, plays a critical role through the increased occurrence of rectangular geometries with specific aspect ratios.
This geometric effect ensures particle mobility and drives fluidization despite identical particle packing fractions.
Future work could explore the introduction of topological and/or non-reciprocal neighbor definitions for the Vicsek interaction, such as those based on Voronoi tessellation, in point particles~\cite{chate_2010,nakamura2025}, hard disks, and binary mixture systems.

\section*{Acknowledgments}
\begin{acknowledgments}
The authors are grateful to Dr.~Daigo Mugita and Mr.~Taisei Nakamura for helpful discussions. M.I.
was supported by JSPS KAKENHI Grant Number 23K03246. Part of the computations were performed using the facilities of the Supercomputer Center, ISSP, Univ.
of Tokyo.
\end{acknowledgments}

\appendix
\section{Model Description}

Active matter systems can be classified by particle type, repulsive interactions, self-propulsion mechanism, energy injection, and conservation laws. We summarize our model's characteristics below.

\noindent
\textit{Excluded volume} --- We employ hard disks to incorporate excluded volume effects, unlike point particles used in the original Vicsek model. While soft-core potentials such as Weeks-Chandler-Andersen (WCA) produce gradual repulsion during overlap, hard disks interact only through instantaneous collisions with no overlap permitted.

\noindent
\textit{Inelasticity} --- Collisions in our hard disk model are elastic, conserving both momentum and kinetic energy. In contrast, granular gas models employ inelastic collisions that dissipate kinetic energy while conserving momentum, and some models also violate momentum conservation through frictional forces.

\noindent
\textit{Self-propulsion } --- Various mechanisms introduce self-propulsion: Vicsek-type alignment with angular or vectorial noise, self-propelled forces in active Brownian particles, or shape-induced rotation in experiments with asymmetric particles~\cite{Deseigne_2010,Weber_2014}. We apply Vicsek-type angular noise~\cite{vicsek_1995} to hard disks at periodic intervals $\Delta t^*$.

\noindent
\textit{Energy injection} --- Non-equilibrium steady states are typically maintained through energy injection via boundary vibration, Langevin thermostats, or velocity rescaling. Our model requires no such injection; when the Vicsek interaction is switched off, the system equilibrates to a Maxwell-Boltzmann velocity distribution.

\noindent
\textit{Conservation laws} --- Microscopic conservation of momentum and energy fundamentally determines macroscopic behavior and the equilibrium or non-equilibrium character of a system. The Vicsek interaction violates local momentum conservation, driving the polar order transition.

The original Vicsek model uses point particles without collisions, applies alignment with angular noise, and constrains each particle to move at a fixed speed without momentum conservation~\cite{vicsek_1995}. Experimental vibrated polar disk systems~\cite{Deseigne_2010,Weber_2014} employ hard disks with inelastic collisions, achieve self-propulsion through shape asymmetry, inject energy via vibration, and violate both conservation laws. Our model uses hard disks with elastic collisions and Vicsek-type alignment at fixed intervals, requiring no external energy injection and conserving both momentum and energy except through the alignment interaction. Thus, only the Vicsek interaction with frequency parameter $\Delta t^*$ introduces non-equilibrium effects into an otherwise equilibrium hard-disk system.

In time-step-driven molecular dynamics with soft-core potentials, the integration step $\delta t$ sets the minimum timescale for Vicsek interactions. This creates ambiguity: alignment events may occur during particle overlap, preventing clean separation of collision and alignment dynamics and introducing $\delta t$-dependence. Hard-disk systems avoid this problem. With no inter-particle potential energy, collisions occur instantaneously, temperature is defined solely by velocity fluctuations, and trajectories are temperature-independent. Near the solid-liquid transition, soft-core systems exhibit complex interplay between potential energy and temperature, requiring both as control parameters. Hard-disk systems are governed by a single parameter: the packing fraction. In event-driven molecular dynamics, Vicsek interactions are introduced as discrete events at arbitrary times, independent of collisions. Both event types are processed sequentially on a continuous timeline. We introduce alignment events at fixed intervals $\Delta t^*$, providing a clear parameter controlling the Vicsek interaction frequency.

\section{Details of the EDMD simulation scheme with Vicsek interaction}
\label{sec:appendix_edmd}

In this study, we employed a hybrid simulation model that incorporates Vicsek-type alignment interactions into a hard disk system governed by Event-Driven Molecular Dynamics (EDMD).
Unlike time-step-driven simulations that integrate equations of motion at fixed small discrete time steps, EDMD advances the system analytically from one discrete event to the next.
This approach is highly efficient for hard-core potentials where interactions occur instantaneously.
Since periodic boundary conditions are applied, there are no hard walls, and disks wrap around the simulation box boundaries.
The events in our system are classified into two main types:

\noindent
\textbf{Elastic Collision} --- A local elastic collision between two parirs of hard disks $i$ and $j$.

\noindent
\textbf{Vicsek Update} --- A global event occurring at fixed intervals $\Delta t$ where the directions of all disks are updated.

The simulation proceeds according to the following algorithm:

\begin{enumerate}
    \item \textbf{Initialization}:
    Set the initial positions $\bm{r}_i$ and velocities $\bm{v}_i$ for all disks at time $t=0$.
    Compute the scheduled time for the next Vicsek update, $t_{\rm VM} = \Delta t$.
    \item \textbf{Event Scheduling}:
    For every pair of disks $i$ and $j$, calculate the time $\Delta t_{ij}$ until their next elastic collision.
    The collision condition is given by $|\bm{r}_{ij} + \bm{v}_{ij} t|^2 = (2\sigma)^2$, where $\bm{r}_{ij} = \bm{r}_i - \bm{r}_j$ and $\bm{v}_{ij} = \bm{v}_i - \bm{v}_j$, considering the minimum image convention for periodic boundaries.
    These event times are stored in an event calendar (e.g., a priority queue).
    \item \textbf{Next Event Selection}:
    Identify the minimum time interval to the next event, $\Delta t_{\min} = \min(\{t_{ij}\}, t_{\rm VM} - t)$.
    \item \textbf{Streaming}:
    Advance the positions of all disks by free streaming: $\bm{r}_i(t + \Delta t_{\min}) = \bm{r}_i(t) + \bm{v}_i(t) \Delta t_{\min}$.
    Periodic boundary conditions are applied to the positions.
    Update the system time: $t \leftarrow t + \Delta t_{\min}$.
    \item \textbf{Event Processing}:
    Execute the event associated with $\Delta t_{\min}$:
    \begin{itemize}
        \item \textit{If Elastic Collision}: Update the velocities of the colliding pair $(i, j)$ according to the laws of conservation of momentum and energy:
        \begin{equation}
        \bm{v}'_i = \bm{v}_i - \frac{\bm{v}_{ij} \cdot \bm{r}_{ij}}{|\bm{r}_{ij}|^2} \bm{r}_{ij}, \quad
        \bm{v}'_j = \bm{v}_j + \frac{\bm{v}_{ij} \cdot \bm{r}_{ij}}{|\bm{r}_{ij}|^2} \bm{r}_{ij}.
        \end{equation}
        \item \textit{If Vicsek Update}: Update the direction $\Theta_i$ of every disks in the system according to Eqs.~(\ref{eqn:vicsek1}) and (\ref{eqn:vicsek2}).
        The speed $v_i = |\bm{v}_i|$ remains unchanged (inherited from the inertial flight), but the velocity vector is reoriented.
        Update the next Vicsek schedule: $t_{\rm VM} \leftarrow t_{\rm VM} + \Delta t$.
    \end{itemize}

    \item \textbf{Recalculation}:
    Recalculate future collision times for the disks involved in the processed event.
    If a Vicsek update occurred, the collision times for all disks must be recalculated since all velocity vectors have changed.
    Update the event calendar and return to Step 3.
\end{enumerate}

\begin{algorithm}[h]
\caption{Simulation Algorithm} 
\SetAlgoLined 
\KwIn{Initial positions $\{\bm{r}_i\}$, velocities $\{\bm{v}_i\}$, update interval $\Delta t$, total simulation time $T$.}
\KwOut{System trajectories and order parameters.}

\BlankLine
\tcp{Initialization}
$t \leftarrow 0$\;
$t_{\text{VM}} \leftarrow \Delta t$ \tcp*[r]{Schedule first Vicsek update}
$Q \leftarrow \text{Initialize event calendar with all predicted collisions } \Delta t_{ij}$\;
\BlankLine
\While{$t < T$}{
    \tcp{Step (2, 3): Next Event Selection }
    $t_{\text{coll}} \leftarrow \text{top}(Q)$\;
    $\Delta t_{\min} \leftarrow \min(t_{\text{coll}} - t, t_{\text{VM}} - t)$\;
    
    \BlankLine
    \tcp{Step (4): Streaming}
    \For{each disk $i$}{
        $\bm{r}_i(t + \Delta t_{\min}) \leftarrow \bm{r}_i(t) + \bm{v}_i(t) \Delta t_{\min}$\;
        Apply periodic boundary conditions to $\bm{r}_i$\;
    }
    $t \leftarrow t + \Delta t_{\min}$\;
    \BlankLine
    \tcp{Step (5): Event Processing}
    \If{$t = t_{\text{coll}}$}{
        \tcp{Elastic Collision between disks $i$ and $j$}
        Update $\bm{v}_i$ and $\bm{v}_j$ using momentum/energy conservation\;
        $Q \leftarrow \text{Recalculate future collisions for disks } i \text{ and } j \text{ }$\;
    }
    \ElseIf{$t = t_{\text{VM}}$}{
        \tcp{Global Vicsek Update}
        \For{each disk $i$}{
            Calculate new direction $\Theta_i(t)$ using alignment rule and noise $\eta$\;
            Update velocity $\bm{v}_i \leftarrow v_i \bm{\hat{e}}_{\Theta_i}$ \tcp*[r]{Speed $v_i$ is conserved }
        }
        $t_{\text{VM}} \leftarrow t + \Delta t$\;
        $Q \leftarrow \text{Clear and recalculate ALL collision times }$\;
    }
}
\end{algorithm}

\bibliographystyle{jpsj}
\bibliography{apssamp}

\end{document}